\documentclass[twocolumn,aps,prb,showpacs,amsmath,superscriptaddress]{revtex4-1}
\usepackage{graphicx}
\usepackage{epstopdf}
\usepackage{color}
\begin{document}

\title{Fulde-Ferrell-Larkin-Ovchinnikov pairing states of a polarized dipolar Fermi gas trapped in a one-dimensional optical lattice}
\author{Xingbo Wei}
\affiliation{Department of Physics, Zhejiang Normal University, Jinhua 321004, China}
\author{Chao Gao}
\affiliation{Department of Physics, Zhejiang Normal University, Jinhua 321004, China}
\author{Reza Asgari}
\affiliation{School of Physics, Institute for Research in Fundamental Sciences (IPM), Tehran 19395-5531, Iran}
\affiliation{School of Nano Science, Institute for Research in Fundamental Sciences (IPM), Tehran 19395-5531, Iran}
\author{Pei Wang}
\affiliation{Department of Physics, Zhejiang Normal University, Jinhua 321004, China}
\author{Gao Xianlong}
\affiliation{Department of Physics, Zhejiang Normal University, Jinhua 321004, China}
\date{\today}
\begin{abstract}
We study the interplay between the long- and short-range interaction of a one-dimensional optical lattice system of two-component dipolar fermions by using the density matrix renormalization group method. The atomic density profile, pairing-pairing correlation function, and the compressibility are calculated in the ground state, from which we identify the parameter region of the Fulde-Ferrell-Larkin-Ovchinnikov (FFLO) pairing state, half-metal (HM) state, FFLO-HM state, and the normal polarized state, and thus the phase diagram in the coordinates of the long- and short-range interaction strength. The effect of the long-range dipolar interaction on the FFLO state is discussed in details. We find that the long-range part of the dipole-dipole interaction does not sweep away the FFLO superconducting region that is driven by the short-range interaction in the Hubbard model, and thus the FFLO state survives in the wide parameter space of the long-range interaction, polarization, and filling.
\end{abstract}

\maketitle

\section{Introduction}
Cold atoms are becoming ideal test beds for performing experiments with fine controllability which makes quantum simulations of many-body systems with ease~\cite{reviewpaper}. The tunability of a wide range of system parameters makes cold atoms a suitable platform which goes beyond the capability of condensed matter physics. In recent years, great advances~\cite{Lahaye,Nik,YiYou,Dell,Lee,Yan,Hazzard,Paz2016,Park-fermiondipolar} have been made in polar molecules by the stimulated Raman adiabatic passage technique, dipolar atomic systems with strong long-range dipole-dipole interactions (DDIs), or dilute dipolar molecules in an optical lattice.

Dipolar quantum gases, in stark contrast to dilute gases with isotropic and extremely short-range contact inter-particle interactions, offer fascinating prospects of exploring many-body novel quantum phases with atomic interactions that are spatially anisotropic and both of long- and short-range interactions. Experimentally the magnitude and the sign of the short-range $s$-wave scattering length characterizing the contact interactions can be tuned by using Fano-Feshbach resonances~\cite{Inouye}, while the dipole-dipole long-range interaction can be adjusted by changing the angle between the lattice orientation and the polarization direction of the dipoles~\cite{Lahaye,Nik}.

The important role of the repulsive long-range interaction in the formation of the Wigner crystal has been analyzed explicitly in the systems both with~\cite{Xuzhihao} and without~\cite{Wangjingjing} optical lattices by studying the density profile and the density-density correlation functions.
The ground-state phase diagram of a Fermi system with both on-site contact and dipole-dipole interactions in a one-dimensional (1D) optical lattice has been studied with the density matrix renormalization group (DMRG) method~\cite{Mosadeq}. In the weak coupling regime, the spin-density wave, charge-density wave, and singlet and triplet superfluidity phases are found; while in the strong coupling regime, bond order wave and phase separation phases are obtained.
The nontrivial Luttinger liquid behaviors of one-dimensional dipolar bosons~\cite{Citro,Guan} or fermions ~\cite{Mosadeq,Silva2013} have been investigated in a wide range of intermediate densities.
In some solid-state materials simulated by the extended Hubbard model, the phase diagram has been identified by using bosonization, DMRG, or exact diagonalization, in which the nearest-neighbour interaction is included~\cite{EHM-Linhaiqing}.

In this paper, we study the phase diagram and ground-state properties of the 1D polarized dipolar fermions by employing the numerically accurate DMRG method~\cite{DMRG}. We focus on the finite-momentum Fulde-Ferrell-Larkin-Ovchinnikov (FFLO)~\cite{FFLO,FFLO-theory,FFLO-num,AHAI} pairing states which originate from the long-range nature of the dipole-dipole interaction. The phase diagram in the coordinates of the long- and short-range interaction strength is identified for a fixed polarization of a quarter-filling (one fermion every two sites, $n=1/2$). Different phases are also observed, including the half-metal (HM) state, FFLO-HM state, and the normal polarized state. The results show that the FFLO state can survive in the large parameter space of long-range interaction, polarization, and filling.

This paper is organized as follows. In Sec.~\ref{sec:model}, we introduce the 1D dipolar lattice fermionic model and explain the protocol that we used in the DMRG calculation. In Sec.~\ref{sec:phase}, the phase diagram is presented according to our analysis of the atomic density profile and pair-pair correlation functions. Section~\ref{sec:con} is devoted to conclusions.

\section{Model and method}
\label{sec:model}
In this study we investigate the FFLO states of a 1D polarized dipolar Fermi gas in an optical lattice, which is described by the following Hamiltonian within a tight-binding and a single-band approximation, with both long- and short-range interactions,
\begin{equation}\label{hubbard}
\hat{H}=-t\sum_{i,\sigma}(\hat{c}_{i\sigma}^\dag\hat{c}_{i+1\sigma}+\text{H.c.})+U\sum_{i}\hat{n}_{i\uparrow}\hat{n}_{i\downarrow}+\frac{g}{2}\sum_{i\neq{j}
}\frac{\hat{n}_{i}\hat{n}_{j}}{|i-j|^3}.
\end{equation}
Here $t$ is the nearest-neighbor hopping amplitude, $\sigma=\uparrow (\downarrow)$ is the pseudospin index, $\hat{c}_{i\sigma}^\dag (\hat{c}_{i\sigma}$) is the creation (annihilation) operator of a fermion at site $i$ with pseudospin $\sigma$, and $U$ and $g$ are the strength of the on-site short-range contact and long-range dipolar interactions, respectively.
$U$ and $g$ can be positive (repulsive interaction) or negative (attractive interaction).
We denote the particle density of the system as $n=(N_{\uparrow}+N_{\downarrow})/L$, where $N_{\sigma}$ is the number of atoms with pseudospin $\sigma$ and $L$ is the length of the chain.
Furthermore, $\hat{n}_{i}$ is the occupation number at site $i$ with $\hat{n}_{i}=\sum_\sigma\hat{n}_{i\sigma}=\hat{n}_{i\uparrow}+\hat{n}_{i\downarrow}$. In this study, we are interested in the polarized system of quarter filling ($n=1/2$). For convenience, the energy unit is chosen to be $t=1$.

Without the long-range interactions ($g=0$), the Hamiltonian reduces to the polarized Fermi-Hubbard model, where the FFLO exists in the polarized case with attractive interaction~\cite{FFLO-ana,FFLO-num}.
In the case of strongly repulsive on-site short- ($U>0$) and dipolar long-range ($g>0$) interactions,
both the density distribution and the static structure factor
display clear signatures of the Wigner crystal phase~\cite{Xuzhihao,Silva2013}.
In this study, we will focus on the pairing nature of this model with attractive on-site short-range ($U<0$) interactions, especially on the parameter space for the existence of the FFLO pairing induced by the interactions of long-range nature.

We investigate the ground states of the dipolar system described by Eq.~(\ref{hubbard}) via the DMRG algorithm under the open boundary condition and study the phase diagram at quarter fillings ($n=1/2$) as a function of $U$ and $g$. In order to ensure the numerical precision, we keep the number of states to be 400 per block and use 40 sweeps resulting in an error from the total weight of the discarded states to be less than $10^{-10}$. When necessary, we increase the sweeps and the states kept to ensure the accuracy. Without specification, we fix $L=40$, $N_{\uparrow}=15$, and $N_{\downarrow}=5$ in our calculations. Some of the calculations are extended to a larger size to eliminate the finite-size effect.
\begin{figure}[htbp]
\begin{center}
\centering
\includegraphics[width=0.5\textwidth]{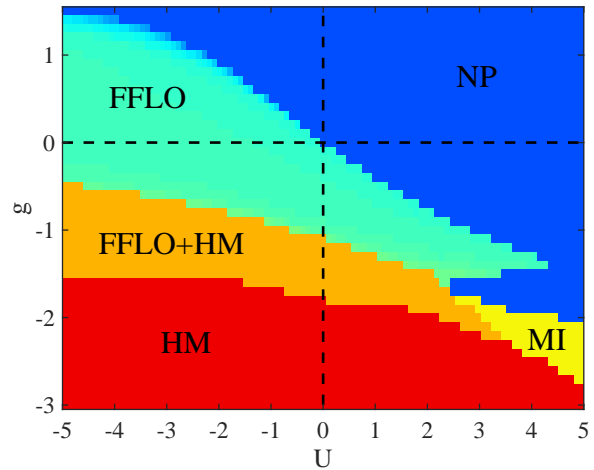}
\end{center}
\caption{(color online) Ground-state phase diagram of a 1D polarized dipolar Fermi gas in an optical lattice obtained by the DMRG method as functions of the long-range interaction strength $g$ and the on-site interaction strength $U$ (in units of $t$). Other parameters we used are $L=40$, $N_\uparrow=15$, and $N_\downarrow=5$.
Five different phases are presented: a normal polarized (NP) phase, an FFLO phase, a half-metal (HM) phase, a Mott-insulating (Mott) phase, and a coexisted FFLO and HM (FFLO-HM) phase. The horizontal and vertical dashed lines are for $g=0$ and $U=0$, respectively. }
\label{Fig1}
\end{figure}

\section{Phase diagram and numerical results}
\label{sec:phase}
To characterize the phases and find the phase boundaries, we study the pair-pair correlation functions defined as,
\begin{equation}\label{pairpair}
G(i,j)=\langle\hat{\Delta}_i^\dag\hat{\Delta}_j\rangle,
\end{equation}
and its Fourier transformations given by,
\begin{equation}\label{pair}
G{_{\text{pair}}} (k)=\Bigr\vert\frac{1}{2L}\sum_{i,j}G(i,j)e^{I(i-j)k}\Bigr\vert,
\end{equation}
where $\hat{\Delta}_i$ is the pair gap operator, defined as
\begin{equation}\label{delta}
\hat{\Delta}_i=\hat{c}_{i\uparrow}\hat{c}_{i\downarrow}.
\end{equation}
Due to the well-known Mermin-Wagner theorem (also known as the Mermin-Wagner-Hohenberg theorem or Coleman theorem),
there is no true long-range order in a 1D system, and as a result, the ground-state expectation value of the pairing gap operator $\hat{\Delta}_i$ is zero~\cite{Giamarchi}. However, the pair-pair correlation functions of the Cooper pair operator have a power-law decay at large distances,
$G(i,j)\sim \vert i-j\vert^{-1/{\kappa_\rho}}$,
where $\kappa_\rho$ is an interaction-dependent dimensionless parameter~\cite{Rizzithesis}.
In the case of unbalanced pseudospin population, the existence of two distinct Fermi surfaces will lead to the creation of pairs with a nonzero total finite momentum. Such a phase, where the pair-pair correlation functions acquire a
spatially dependent non-uniform oscillatory character, is known as an FFLO state~\cite{reviewpaper,Mosadeq,Giamarchi}.

Our main results are summarized in the ground-state phase diagram shown in Fig.~\ref{Fig1}. Five different phases are obtained: a normal polarized (NP) phase, an FFLO phase, an HM phase, a Mott-insulating (MI) phase, and a coexisted FFLO and HM (FFLO-HM) phase. The detailed analysis will be addressed with calculations for the density profile, pair-pair correlation functions, and local compressibility.
\begin{figure}[htbp]
\begin{center}
\centering
\includegraphics[width=0.45\textwidth]{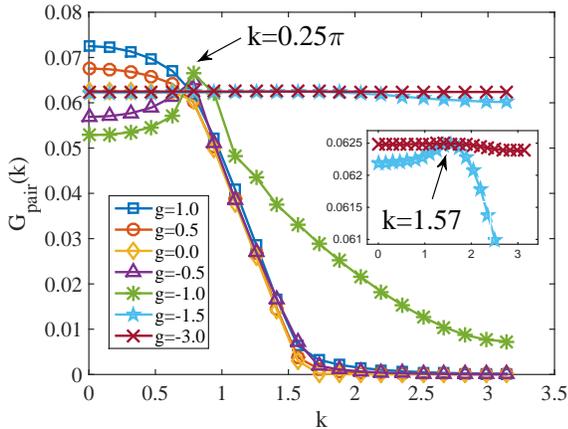}
\end{center}
\caption{(color online) Pair momentum distributions for the vanishing on-site interaction ($U=0$) as a function of $k$ for the different amplitude of the long-range interaction $g$ at $g=1.0, 0.5, 0, -0.5, -1.0, -1.5, -3.0$. The pair momentum distributions show a peak around $k=0.25\pi$ for $g=-0.5, -1.0$. The inset shows an enlargement of the region around the finite-momentum peak $k=1.57$ for $g=-1.5$ and flat distribution for $g=-3.0$. The disappearance of the finite peak indicates the loss of the long-range order pair correlation. }
\label{Fig2}
\end{figure}

For a polarized system with only attractive on-site interactions ({\it i.e.}, $g=0$), the system opens a gap in the spin sector which induces an exponential decay of spin correlations, while singlet superconducting and charge-density wave correlations have a power-law decay.
In such a phase, Cooper pairing yields a spatially dependent superconducting order parameter at a finite momentum ($k\ne 0$) equal to the distance between the two distinct Fermi surfaces $k=k_{F\uparrow}-k_{F\downarrow}$,
where $k_{F\sigma}$ is the Fermi vector of the $\sigma$ fermions.
Thus, a peak appears in the pair momentum distribution $G_{\text{pair}}(k)$ at $k\ne 0$,
which serves as an order parameter of FFLO states,
and indicates a long-range order pair correlation in the system for finite polarization~\cite{FFLO-num,AHAI,FFLO-theory}. The peak of the pair momentum distribution is predicted in a homogeneous system as $k=\pi (N_{\uparrow}-N_{\downarrow})/L$~\cite{FFLO-theory}.

\subsection{Vanishing on-site interaction}
We first discuss the possible finite-momentum FFLO pairing states with only repulsive or attractive long-range dipolar interactions ($U=0$).
The results are shown in Fig.~\ref{Fig2}.
For the repulsive long-range interaction ($g=1.0, 0.5$), there is no finite-momentum peak in $G_{\text{pair}}(k)$.
The system is in the normal polarized phase.
For the attractive long-range interaction with relatively weak strength ($g=-0.5, -1.0$),
an obvious peak appears at $k=0.25\pi\approx 0.785$,
which is in accordance with that in the homogeneous case
$k=\pi(N_{\uparrow}-N_{\downarrow})/L=\pi n p$, and is a signature of the FFLO phase.
While the attractive dipole-dipole interaction continues to increase, the peak (shifting to $k=1.57$)
becomes vanishingly small at $g=-1.5$.
By further increasing the attractive interaction ($g=-3.0$),
 $G_{\text{pair}}(k)$ becomes flat, signalling the disappearance of the FFLO state and indicating an HM phase, which will be clear in the following analysis.
\begin{figure}[htbp]
\begin{center}
\centering
\includegraphics[width=0.45\textwidth]{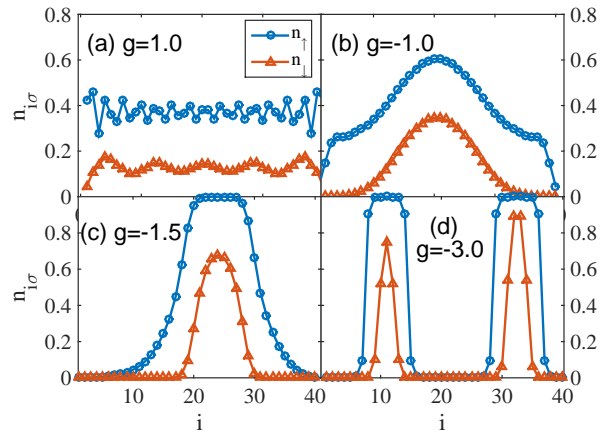}
\end{center}
\caption{
Density distribution for spin-up and spin-down atoms as a function of the lattice site for
 (a) $g=1.0$, (b) $g=-1.0$, (c) $g=-1.5$, and (d) $g=-3.0$ in the case of vanishing on-site interaction ($U=0$).
 The other parameters used are the same as those in Fig.~\ref{Fig2}. }
\label{Fig3}
\end{figure}
\begin{figure}[htbp]
\begin{center}
\centering
\includegraphics[width=0.45\textwidth]{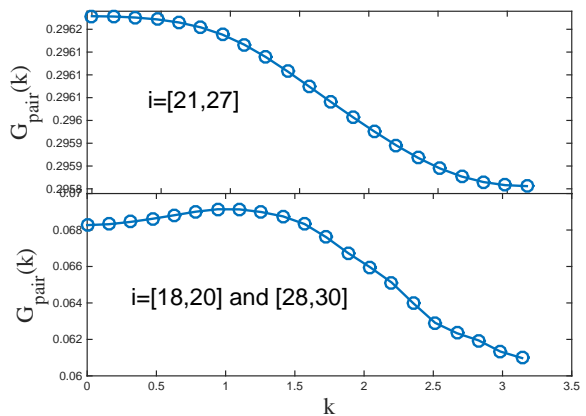}
\end{center}
\caption{(color online) Pair momentum distributions in an HM region ($i\in[21,27]$ ) (top panel) and a fully polarized region ($i\in[18,20]~ \cup \in[28,30]$) for $U=0$ and $g=-1.5$ (bottom panel).}
\label{Fig4}
\end{figure}

The atomic density distributions of different pseudospins for $g=1.0, -1.0, -1.5, -3.0$ are shown in Fig.~\ref{Fig3}.
In Fig.~\ref{Fig3}(a)-~\ref{Fig3}(b), we show the normal polarized phase and the FFLO phase, respectively, where both $n_{i\uparrow}$ and $n_{i\downarrow}$ $\in [0,1)$.
In the FFLO state, normally there exists a clear fluctuation in the density distribution~\cite{FFLO-num,AHAI,FFLO-theory}. However, owing to the attractive long-range nature of the dipole-dipole interaction, the fluctuation of density distribution is not obvious.

While for $g=-1.5$ as shown in Fig.~\ref{Fig3}(c), it is evident that the wings of the density profile are polarized, and the spin-up atoms fill the core of the lattice with a plateau of local density $n_{i\uparrow}=1$, corresponding to an HM state ($n_{i\uparrow}=1$, $0<n_{i\downarrow}<1$)~\cite{Iskin}; that is, it is a metal phase for spin-down atoms and an insulating phase for spin-up. However, as depicted in Fig.~\ref{Fig2}, a very small finite momentum peak in $G_{\text{pair}}(k)$ exists, indicating an FFLO phase.
Further analysis shows that the particles at wings are of FFLO character and the particles at the core form an HM state.
By separating Fig.~\ref{Fig3}(c) into two parts, as is shown in Fig.~\ref{Fig4}, we plot the pair momentum distributions in the HM region
(for $i\in[21,27]$) in the top panel, and the fully polarized region (for $i\in[18,20]$ and $i\in[28,30]$) in the bottom panel.
In the HM region, the pair momentum distribution decreases slowly as $k$ changes without a finite momentum peak,
while in the fully polarized region, there is an obvious peak at $k (\neq 0.25\pi)$~\cite{Luscher}. By further studying larger systems, we confirm that the long-range interaction induced phase separation causes the localization of the particles and the weak FFLO phase by the polarized edge persists.
As a result, what we obtained is a coexisted FFLO and HM (FFLO-HM) phase.

For even stronger attractive dipole-dipole interactions,
as is shown in Fig.~\ref{Fig3} (d) for $g=-3.0$,
the spin-up density almost becomes either $n_{i\uparrow}=1$ or $n_{i\uparrow}=0$.
The correlation between different lattices vanishes $G(i\neq{j})=0$ and thus, the pair momentum distributions are simplified into
$G_{\text{pair}}(k)=|\sum_{i}G(i,i)/{2L}|=|\sum_{i}(-n_{i\downarrow})/{2L}|=N_{\downarrow}/{2L}$.
 As a result, for $g=-3.0$, $N_{\downarrow}=5$, $L=40$, $G_{\text{pair}}(k)=0.0625$, as can be seen in the inset of Fig.~\ref{Fig2}.

In principle, a pure dipolar system with vanishing on-site interactions can be realized in experiments by means of a magnetic Feshbach resonance, which allows reducing the strength of the short-range interactions and thus enhancing the dipolar effects.
\begin{figure}[htbp]
\begin{center}
\centering
\includegraphics[width=0.45\textwidth]{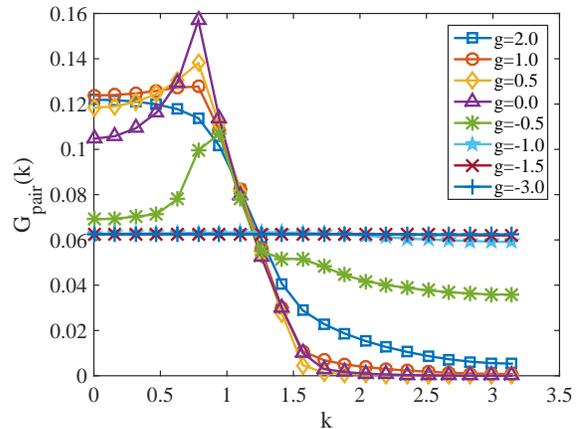}
\end{center}
\caption{(color online) The pair momentum distributions in the case of finite attractive on-site interaction ($U=-4$) as a function of $k$ for the different amplitude of the long-range interaction $g$ at $g=2.0,1.0, 0.5, 0, -0.5, -1.0, -1.5, -3.0$. The pair momentum distributions show a peak for $g=1.0, 0.5, 0, -0.5$. For $g=-1.0, -1.5$, the almost flat curves indicate a coexisted FFLO-HM phase. For $g=-3.0$, the pair momentum distributions become a straight line indicating an HM phase.}
\label{Fig5}
\end{figure}

\subsection{Attractive on-site interaction}
With the on-site interaction non-vanishing,
we now explore the interplay of the on-site interaction and the long-range interaction on the FFLO states.
First, we discuss the $U<0$ case.
Without long-range interactions ($g=0$), the FFLO phase has been analyzed~\cite{FFLO-num,Heikkinen}.
We find that, as shown in Fig.~\ref{Fig5} ($U=-4$), in the presence of the repulsive long-range interaction ($g=0.5$ and $g=1.0$),
for the pair momentum distributions, there is still a peak at $k=0.25\pi$, signalling the existence of the FFLO states.
For larger repulsive dipole-dipole interaction such as $g=2.0$, the FFLO state is destroyed, and the system is in the NP phase.
For $g<0$, the attractive dipole-dipole interaction induces the FFLO states as those for $U=0$ in Fig.~\ref{Fig2}.
When $g=-1.0, -1.5$, the almost flat curves indicate a coexisted FFLO-HM phase.
For $g=-3.0$, the pair momentum distributions become a straight line indicating an HM phase.
\begin{figure}[htbp]
\begin{center}
\centering
\includegraphics[width=0.45\textwidth]{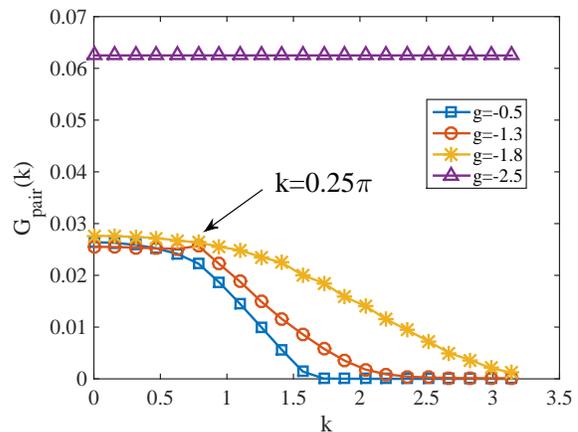}
\end{center}
\caption{(color online) The pair momentum distributions in the case of finite repulsive on-site interaction ($U=3$) as a function of $k$ for the different amplitude of the long-range interaction $g$ at $g=-0.5, -1.3, -1.8, -2.5$ and the system is in a NP, an FFLO, a MI, and an HM phase, respectively. }
\label{Fig6}
\end{figure}

\subsection{Repulsive on-site interaction}
We now explore the interplay of the on-site repulsive interaction and the long-range interaction on the FFLO states.
In Fig.~\ref{Fig6},
we present the pair momentum distributions for different values of $g$ in the situation of the repulsive on-site interaction $U=3$.
Comparing to the case of $U=0$ in Fig.~\ref{Fig2}, the finite momentum peak $k \neq 0$ vanishes at much smaller attractive long-range interaction ($g=-0.5$), due to the competition between the repulsive on-site interaction and the attractive dipole-dipole interaction.
With the increase of the attractive dipole-dipole interaction, the FFLO pairing order dominates.
A peak appears at $k=0.25\pi$ (for $g=-1.3$), signalling the existence of the FFLO state.
While further increasing the attractive dipole-dipole interaction, the peak vanishes (for $g=-1.8$), which indicates a Mott-insulating (MI) phase, as will be exploited in details in the following.
And finally $G_{\text{pair}}(k)$ turns into a straight line and the system is in an HM phase (for $g=-2.5$).
\begin{figure}[htbp]
\begin{center}
\centering
\includegraphics[width=0.45\textwidth]{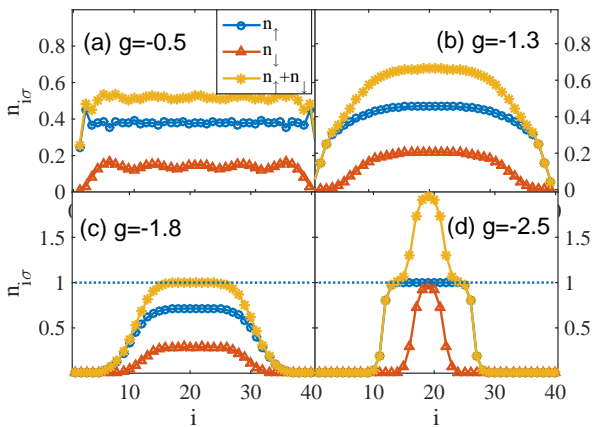}
\end{center}
\caption{(color online) The distribution of individual spin (up, down) density and total density, for $U=3.0$ and different values of $g$. (a) for $g=-0.5$ in a NP phase, (b) for $g=-1.3$ in an FFLO phase, (c) for $g=-1.8$ in a MI phase, and (d) for $g=-2.5$ in an HM phase. The dotted line represents $n_i=1$.}
\label{Fig7}
\end{figure}

\begin{figure}[htbp]
\begin{center}
\centering
\includegraphics[width=0.45\textwidth]{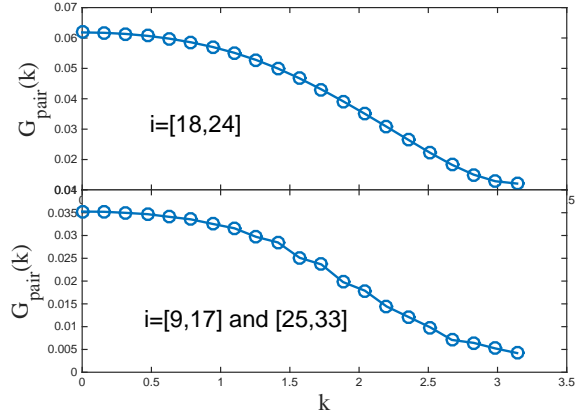}
\end{center}
\caption{(color online) The pair momentum distributions for $U=3.0$ and $g=-1.8$ for two separated regions, where the density profile is plotted in Fig.~\ref{Fig7}(c). Top panel: for a MI region of $i\in[18,24]$, and bottom panel: for a fully polarized region of $i\in[9,17]$ and $i\in[25,33]$.}
\label{Fig8}
\end{figure}

To clarify different phases, we investigate the behavior of the density profile in Fig.~\ref{Fig7} for $U=3.0$.
For the NP phase ($g=-0.5$), the FFLO phase ($g=-1.3$), and the HM phase ($g=-2.5$),
the density profile is similar to those plotted in Fig.~\ref{Fig3},
while for $g=-1.8$, the core of the lattice with a plateau of density $n_i=n_{i\uparrow}+n_{i\downarrow}=1$, corresponding to an MI phase~\cite{Moreno,Machida1,MachidaPRB}. Comparing the Fermi-Hubbard model, we conclude that the MI phase is robust against the long-range attractive interactions.
On the wings, the different pseudospin atoms are fully polarized.
By further investigating the pair momentum distributions of the MI region and the fully polarized region, as is shown in Fig.~\ref{Fig8}, we find that there is no longer a peak at $k\neq0$ in both regions, which means that the fully polarized region is a NP region.
Since the polarized edge is a normal phase, the MI phase we studied is also a mixed phase, which should be named the mixed NP-MI phase. However, without making confusion, we still call it the MI phase.

\begin{figure}[htbp]
\begin{center}
\centering
\includegraphics[width=0.45\textwidth]{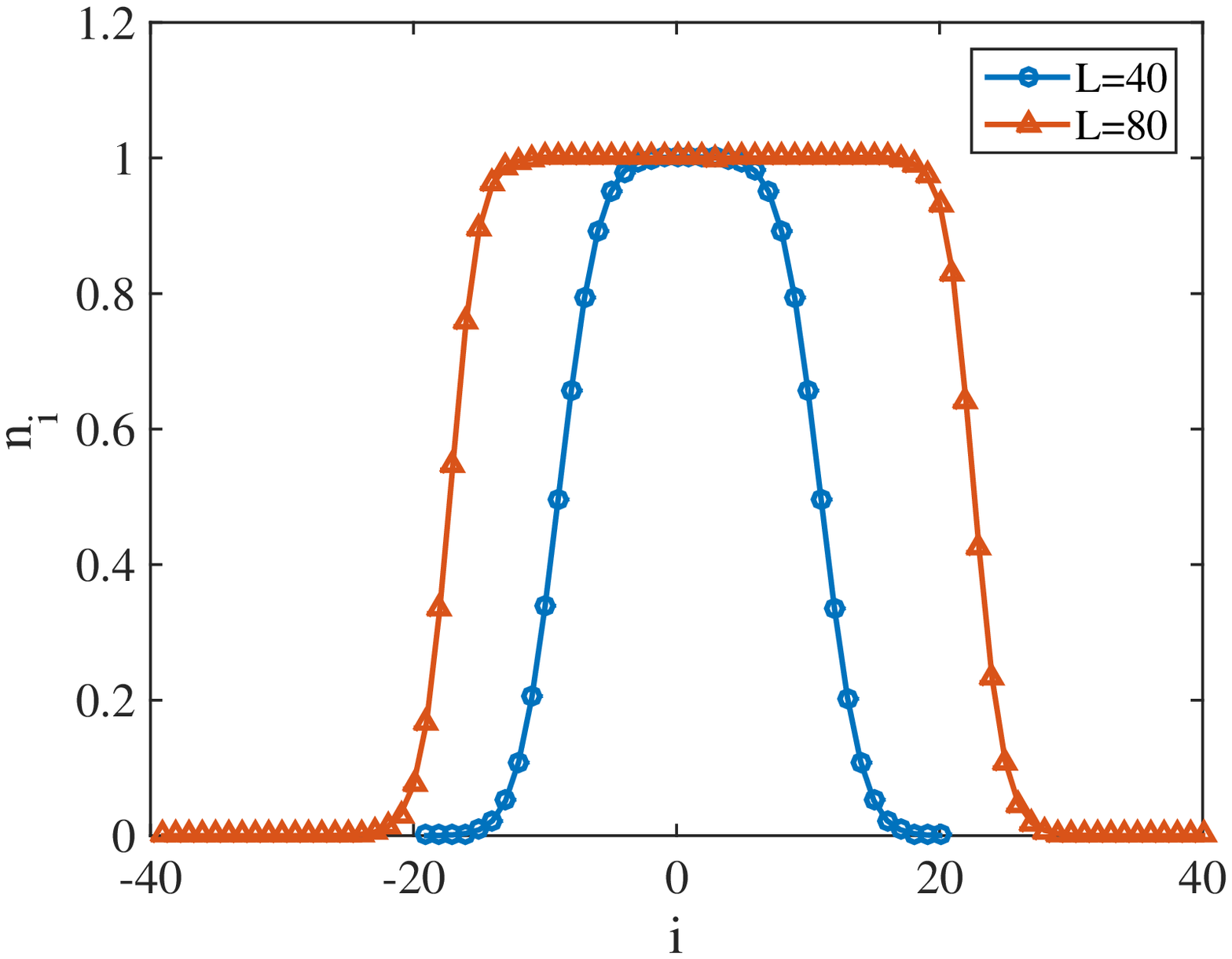}
\includegraphics[width=0.45\textwidth]{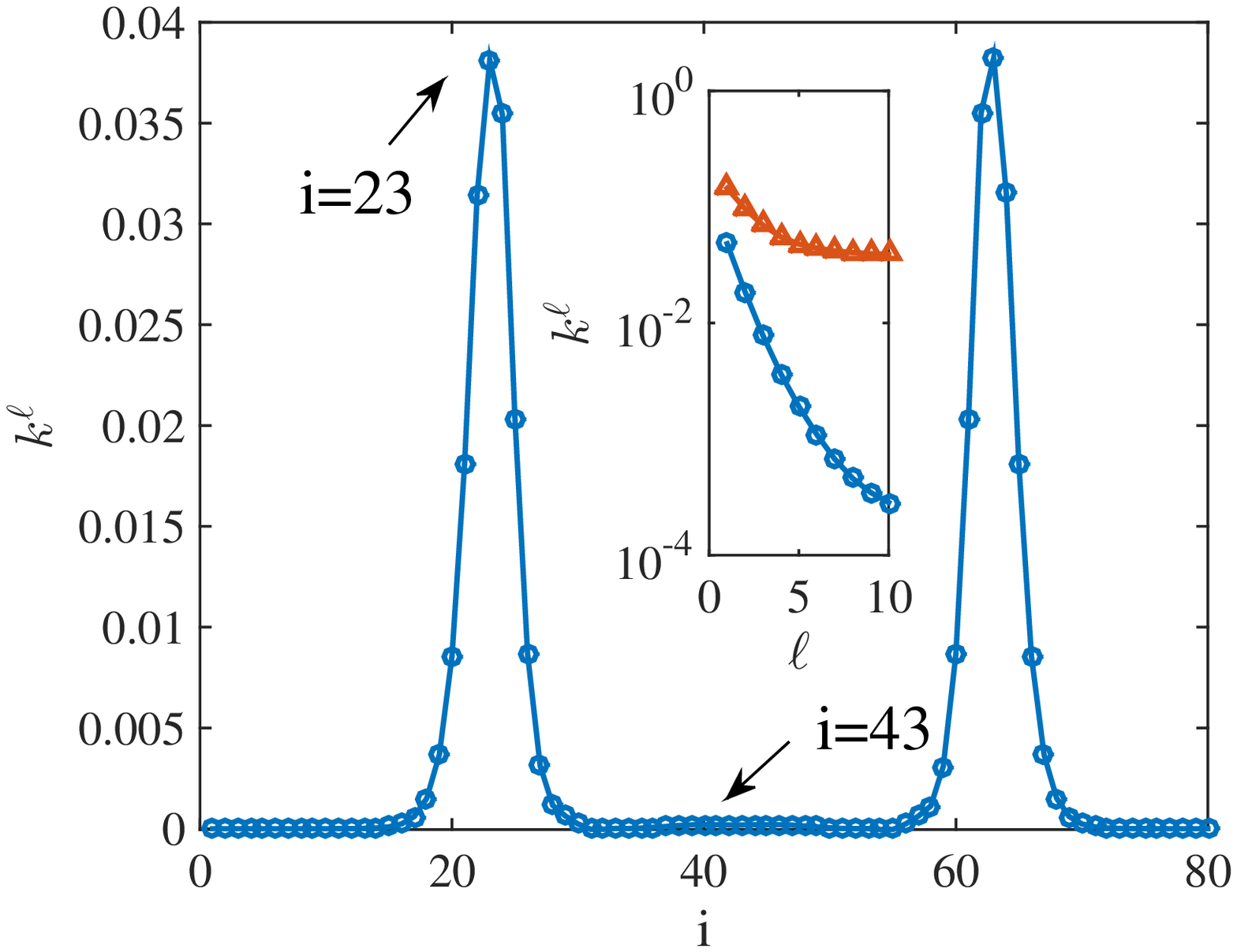}
\end{center}
\caption{(color online) The density distribution (top panel) and its local compressibility (bottom panel) for $L=40$ ($N_{\uparrow}=15$, $N_{\downarrow}=5$), and $L=80$ ($N_{\uparrow}=30$, $N_{\downarrow}=10$). The interaction parameters are $U=3.0$ and $g=-1.8$. Inset: The local compressibility as a function of $\ell$ for $i=43$ ($\bigcirc$) and $i=23$ ($\bigtriangleup$).}
\label{Fig9}
\end{figure}

To further confirm the system with $U=3.0$ and $g=-1.8$ to be in the MI phase, we calculate the local compressibility~\cite{Rigol1,Rigol2}, which is defined as
\begin{equation}\label{kl}
k_i^\ell=\sum_{|j|\leq\ell(U)}(\langle n_in_{i+j}\rangle - \langle n_i\rangle \langle n_{i+j} \rangle),
\end{equation}
where $\ell(U)\simeq{b\xi(U)}$. $\xi(U)$ is the correlation length given by density-density correlation function in the unconfined system at half filling for a given value of $U$. The factor $b$ is chosen to make $k_i^\ell$ qualitatively insensitive to $\xi(U)$ ($\sim{a}$, and $a$ is the lattice constant). From Eq.~(\ref{kl}), we know that if the range of the MI region is less than $\ell(U)$, the local compressibility will be greatly affected. For the system size in Fig.~\ref{Fig7}(c), the range of MI regions is not large enough, so we scale up the system to $L=80$. The density distributions and their local compressibility for $L=40$ and $80$ are shown in Fig.~\ref{Fig9}. In the MI regions, the density distribution shows a plateau with $n_i=1$ and its $k_{\ell}=0$ due to the charge gap present there.
We notice that in the MI phase we studied in the paper (with a quarter filling), the system has the ferromagnetic correlations inside the Mott phase, that is, the spin density $n_{i\uparrow}-n_{i\downarrow}$ is always positive. However, when the polarization becomes smaller, ferromagnetic spin-density wave correlations inside the Mott phase could be antiferromagnetic.
\begin{figure}[htbp]
\begin{center}
\centering
\includegraphics[width=0.45\textwidth]{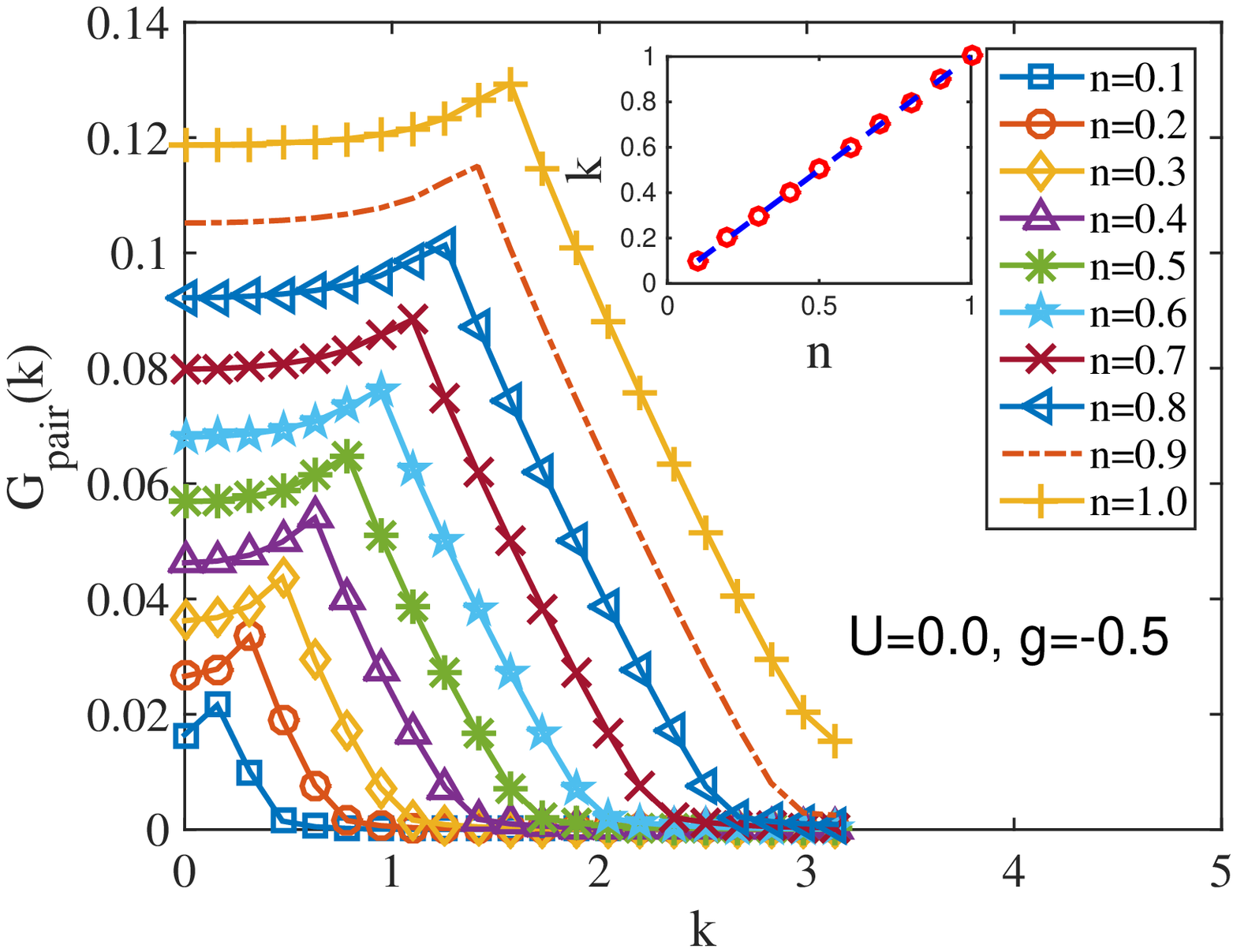}
\includegraphics[width=0.45\textwidth]{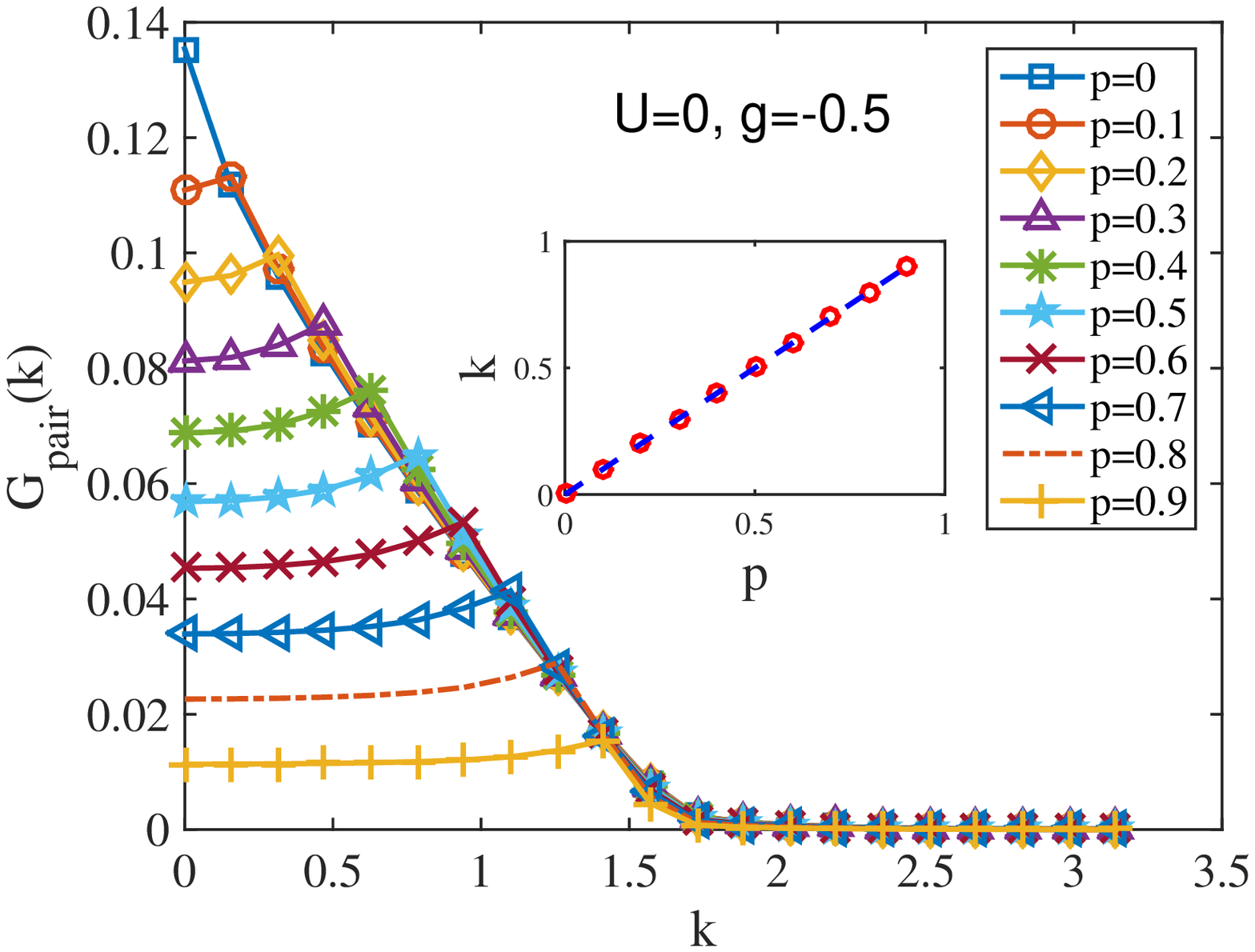}
\end{center}
\caption{(color online) The pair momentum distributions for different fillings (top panel) and polarizations (bottom panel) for $U=0.0$ and $g=-0.5$.
 Inset: The peak of the $G_{\text{pair}}(k)$ versus $k$, which satisfies $k_{\bm FFLO} = \pi n p$. Here $p$ is the polarization of the system.}
\label{Fig10}
\end{figure}

In Fig.~\ref{Fig10}, we study the pair momentum distributions for different polarizations (top panel) and different fillings (bottom panel) with the attractive long-range dipole-dipole interaction at $g=-0.5$ and vanishing on-site interaction $U=0$. Inset shows the peak of the $G_{\text{pair}}(k)$ versus $k$, which satisfies $k_{\bm FFLO} = \pi n p$. Here $p$ is the polarization of the system defined as
$p=(N_\uparrow-N_\downarrow)/N$. Comparing to the polarized attractive Fermi-Hubbard model, where a critical polarization exists below which the FFLO state emerges, for the long-range dipolar system, FFLO survives for the whole range of the polarization~\cite{Franca}.
\begin{figure}[htbp]
\begin{center}
\centering
\includegraphics[width=0.45\textwidth]{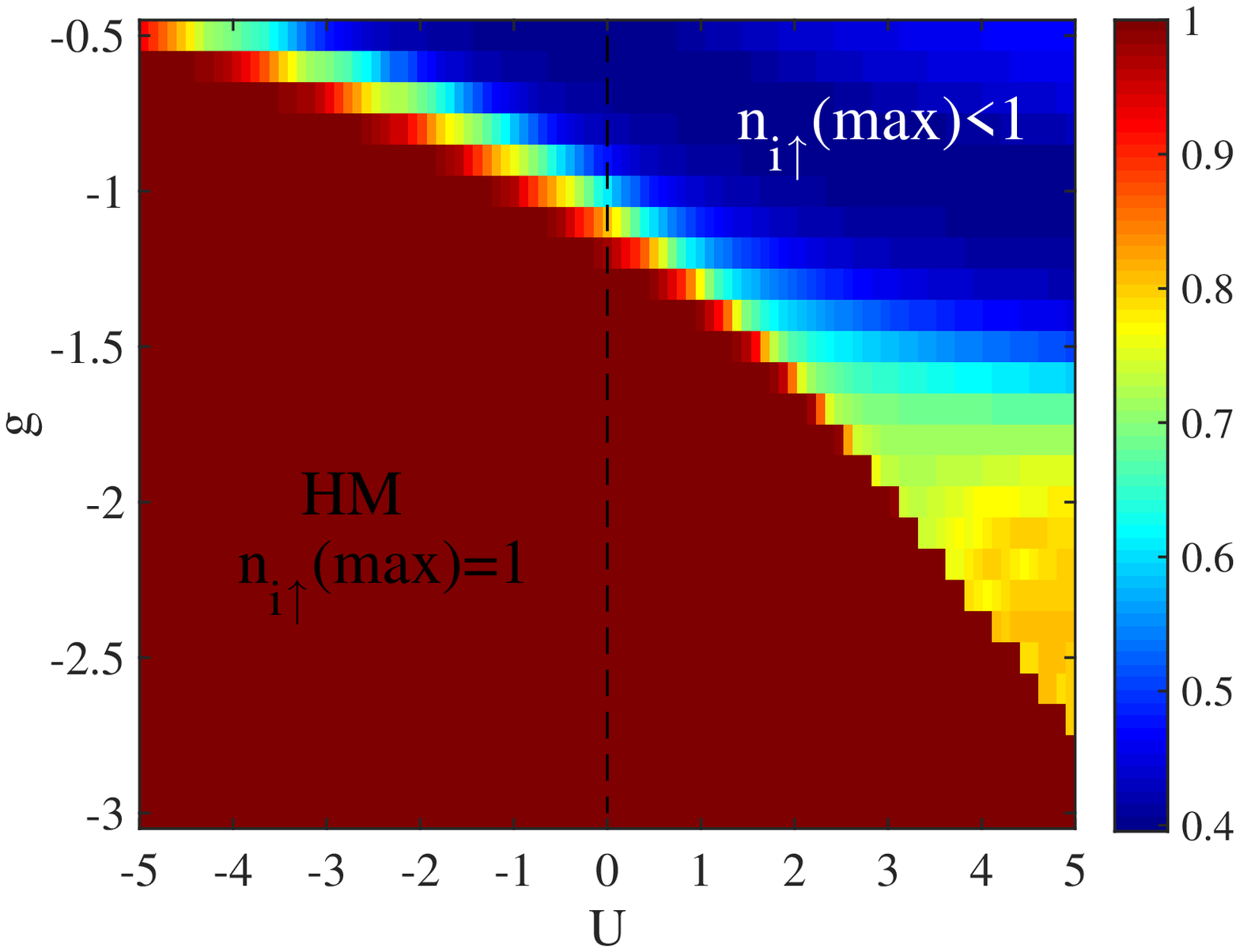}
\includegraphics[width=0.45\textwidth]{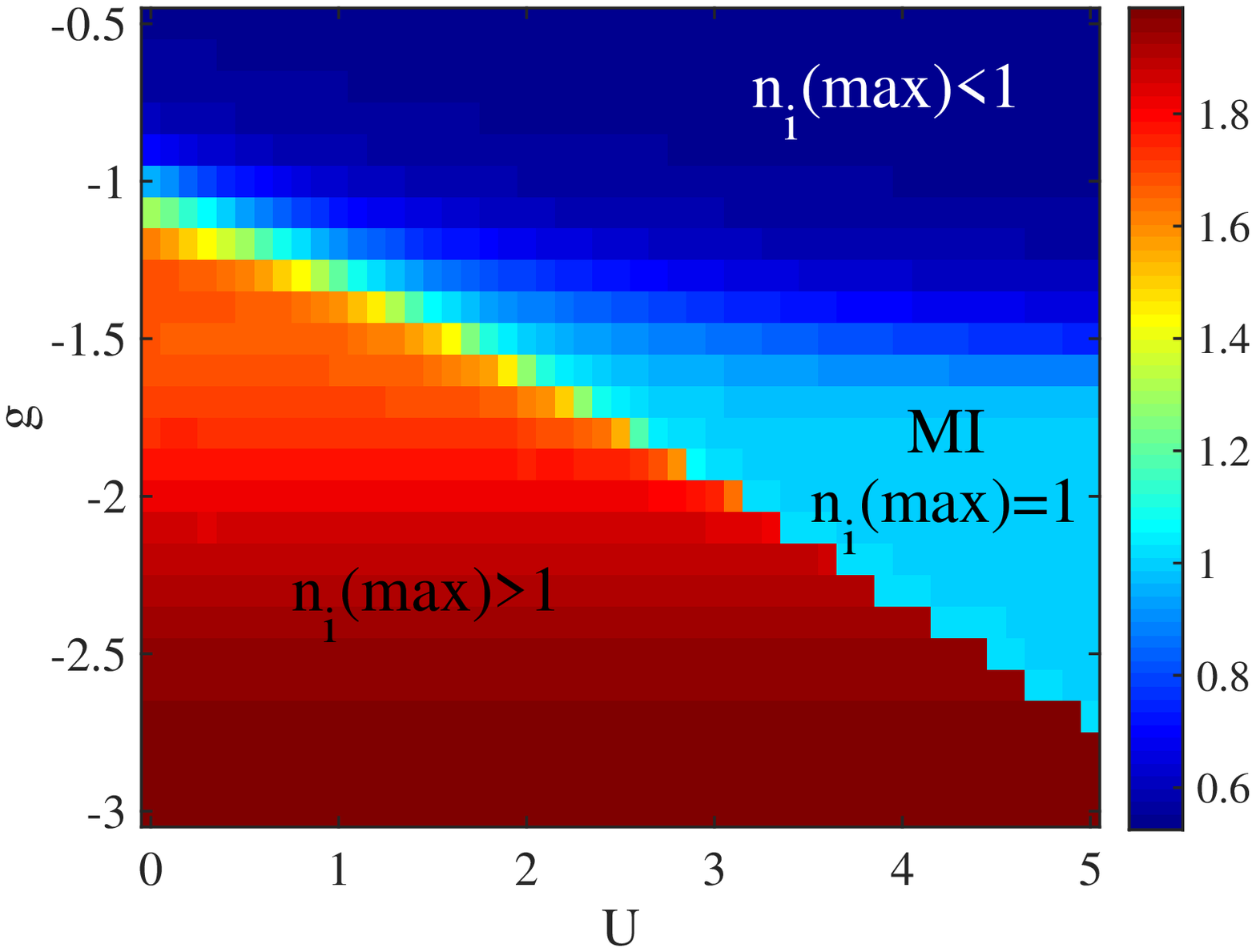}
\includegraphics[width=0.45\textwidth]{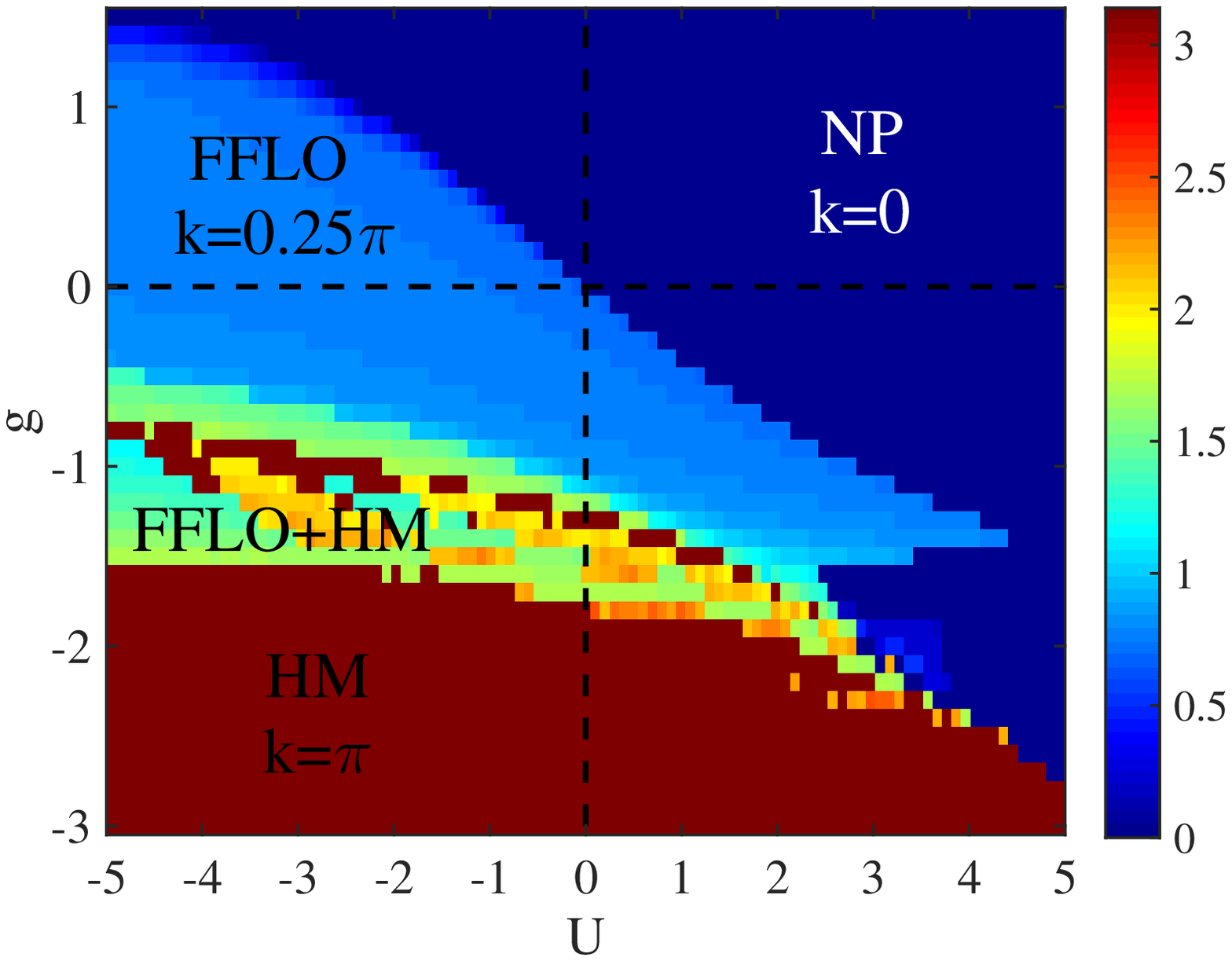}
\end{center}
\caption{(color online) Top panel: the distribution of $n_{i\uparrow}{(max)}$ as a function of $U$ and $g$,
 and the phase boundary between the HM phase and other phases.
  In the HM phase, spin-up atoms fill the core of the lattice with a plateau of unit local density, thus $n_{i\uparrow}{(max)}=1$; while in the other phases $n_{i\uparrow}{(max)}<1$.
Middle panel: the distribution of $n_{i}(max)$ and the phase boundary between the MI, HM, and NP phase.
 The MI phase is determined by $n_{i}{(max)}$ occupying the lattice with unit particle number.
  For the HM phase, $n_{i}{(max)}>1$, while for the NP phase, $n_{i}{(max)}<1$.
Bottom panel: the distribution of the peak  $k_{\text{peak}}$ of the pair distribution $G_{\text{pair}}(k)$,
   and the phase boundary between the FFLO, FFLO-HM, NP phases.
   The pure FFLO phase is distinguished by a finite peak for $G_{\text{pair}}(k)$ at $k=0.25\pi$.
    The FFLO-HM phase has a finite $k$ away from $0.25\pi$.
     The NP phase has a peak at $k=0$.
     Since the HM phase happens with a flat distribution for $G_{\text{pair}}(k)$, we assign $k_{\text{peak}}=\pi$.}
\label{Fig11}
\end{figure}

\subsection{Phase diagram}
Based on the above analysis, in the following, we illustrate in Fig.~\ref{Fig11} by describing how to determine the phase diagram, as in Fig.~\ref{Fig1}.
In Figs.~\ref{Fig1} and~\ref{Fig11}, the values of meshes ($\delta g$, $\delta U$) swept through the figure range are $0.1$ for each axis.

In the top panel, we determine the phase boundary between the HM and other phases by calculating the maximum of $n_{i\uparrow}$: $n_{i\uparrow}{(max)}$.
 In the HM phase, the spin-up atoms fill the core of the lattice with a plateau of unit local density, thus $n_{i\uparrow}{(max)}=1$,
while in the other phases $n_{i\uparrow}{(max)}<1$.
In the middle panel, we distinguish the MI phase by $n_{i}{(max)}$ occupying the lattice with unit particle number.
 For the HM phase, $n_{i}{(max)}>1$, while for the NP phase, $n_{i}{(max)}<1$.
In the bottom panel, the FFLO phase is judged from the pair-pair correlation function with the finite center-of-mass momentum $k$.
 According to the analysis above, we know that the pure FFLO phase is distinguished by a finite peak for $G_{\text{pair}}(k)$ at $k=0.25\pi$, the FFLO-HM phase at finite $k$ away from $0.25\pi$, and the NP phase at $k=0$.
Since the HM phase happens with a flat distribution for $G_{\text{pair}}(k)$, we choose the value at $k=\pi$ for reference. Numerically when $\Delta{G_{\text{pair}}(k)}<0.00001$ at $k=\pi$, we identify that the FFLO phase is completely destroyed.

\section{Conclusions}
\label{sec:con}
In summary, in this work we have investigated the ground state properties of a one-dimensional two-component Fermi gas with population imbalance loaded in an optical lattice with both long-range dipole-dipole and short-range contact interactions.
Making use of the DMRG method, we have studied the interplay of the on-site interaction and the dipole-dipole interaction on the FFLO states.

We have shown that the weak attractive dipole-dipole interaction can lead to the formation of the FFLO states even for the repulsive on-site short range interaction. With the increase of the attractive dipole-dipole interaction, the mixed FFLO-HM phase appears, and finally FFLO states are completely
replaced by the HM phase.
On the other hand, for the attractive on-site interaction, the FFLO phase is robust in the case of repulsive long-range interaction ($g>0$). For much larger $g>0$, the NP phase forms. In the case of the strong repulsive on-site interaction and strong attractive long-range interaction, the Mott phase is favored with unit local density, accompanying zero local compressibility.

Observation of the FFLO phase in a system with dipolar interactions is within the reach of current and future experiments in elongated traps~\cite{RFF,FFLO-exp,Korolyuk}. If ultracold dipolar quantum gases were to be created in quasi-1D tubes, we predict the observation of a finite peak in the pair-pair distribution function as fingerprints of the FFLO behavior. For the influence of the finite temperature, according to the previous research, the exotic FFLO phase survives below one-tenth of Fermi degeneracy temperature for the system without optical lattices and for $T/T_F<0.11$ (with $T_F$ the Fermi temperature), a temperature higher than that of the experiment ($T/T_F>0.11$) for the system in the optical lattice of harmonic confinement~\cite{finte-temp}.

\begin{acknowledgments}
The DMRG calculations were performed using the ALPS libraries.
This work was supported by the NSF of China under Grant Nos. 11774316, 11774315 and 11604300, and the NSF of Zhejiang Province under Grant No. Z15A050001.
\end{acknowledgments}

\end{document}